\documentclass[letterpaper, 10 pt, conference]{ieeeconf}  

\IEEEoverridecommandlockouts                              

\usepackage{epsfig} 
\usepackage{mathptmx} 
\usepackage{times} 
\usepackage{amsmath} 
\usepackage{amssymb}  
\usepackage{graphicx}
\usepackage{svg}
\usepackage{parskip}
\usepackage{float}

\title{\LARGE \bf
Learning Subsystem Dynamics in Nonlinear Systems via Port-Hamiltonian Neural Networks*
}

\author{G.J.E. van Otterdijk$^{1}$, S. Moradi$^{1,2}$, S. Weiland$^{1,2}$, R. T\'{o}th$^{1,2,4}$, N.O. Jaensson$^{2,3}$, M. Schoukens$^{1,2}$%
\thanks{*This work is part of the DAMOCLES research project which received funding from the Eindhoven Artificial Intelligence Systems Institute, as part of the EMDAIR funding program. Furthermore, this work is funded by the European Union (ERC, COMPLETE, 101075836). Views and opinions expressed are however those of the author(s) only and do not necessarily reflect those of the European Union or the European Research Council Executive Agency. Neither the European Union nor the granting authority can be held responsible for them.}%
\thanks{$^{1}$ Control Systems Group, Eindhoven University of Technology, Eindhoven, the Netherlands. $^{2}$ Eindhoven Artificial Intelligence Systems Institute, Eindhoven, the Netherlands. $^{3}$ Processing and Performance of Materials Group, Eindhoven University of Technology, Eindhoven, the Netherlands. $^{4}$ Systems and Control Laboratory, Institute for Computer Science and Control, Budapest, Hungary.}%
\thanks{**E-mail addresses: {\tt\small g.j.e.v.otterdijk@tue.nl,  s.moradi@tue.nl, s.weiland@tue.nl, r.toth@tue.nl. n.o.jaensson@tue.nl, m.schoukens@tue.nl}.}%
}

\begin{document}
\maketitle
\thispagestyle{empty}
\pagestyle{empty}

\begin{abstract}
Port-Hamiltonian neural networks (pHNNs) are emerging as a powerful modeling tool that integrates physical laws with deep learning techniques. While most research has focused on modeling the entire dynamics of interconnected systems, the potential for identifying and modeling individual subsystems while operating as part of a larger system has been overlooked. This study addresses this gap by introducing a novel method for using pHNNs to identify such subsystems based solely on input-output measurements. By utilizing the inherent compositional property of the port-Hamiltonian systems, we developed an algorithm that learns the dynamics of individual subsystems, without requiring direct access to their internal states. On top of that, by choosing an output error (OE) model structure, we have been able to handle measurement noise effectively. The effectiveness of the proposed approach is demonstrated through tests on interconnected systems, including multi-physics scenarios, demonstrating its potential for identifying subsystem dynamics and facilitating their integration into new interconnected models.
\end{abstract}

\section{INTRODUCTION}
Modeling and identifying complex nonlinear dynamical systems present significant challenges in system identification. Selecting the right method to overcome these challenges is non-trivial \cite{SI_methods_roadmap}, as numerous methods have been developed \cite{SI_nonlinear_book}, including various machine learning based algorithms.  Among these, Hamiltonian neural networks (HNNs) have emerged as promising techniques that integrate data-driven modeling with physical principles \cite{HNN_1}. HNNs are based on the concept of energy conservation, where the model structure is based on the Hamiltonian theory. This approach improves the interpretability and robustness of the learned model, especially in long-term simulations, as the dynamics are less prone to divergence with HNNs compared to purely data-driven methods \cite{HNN_2, HNN_3, moradi2023physics}.

Building on HNNs, port-Hamiltonian neural networks (pHNNs) extend the framework to include dissipative elements and external inputs, allowing for the modeling of a wider range of physical systems \cite{ph_identification_1, ph_identification_2}. The port-Hamiltonian approach enables the representation of systems as networks of interconnected subsystems, where ports are used to model flows, such as power, across different elements \cite{ph_theory, ph_book2}. This port-based modeling framework facilitates the simulation of interconnected multiphysics systems, where components from various fields, such as mechanical, electrical, or thermal systems, interact seamlessly \cite{ph_crossdomain}. Several adaptations of pHNNs have been proposed to capture these dynamics using neural networks \cite{ph_identification_3}.

In real-world applications, the ability to model subsystems within the context of interconnected systems is crucial. Capturing the behavior of each subsystem within its operational context enables more accurate simulations and insights into how the overall system responds to varying conditions. Furthermore, a modular approach allows subsystems to be developed, tested, and optimized independently before being integrated into a complete system, which enhances flexibility and scalability. While previous studies have addressed linear interconnected systems \cite{fonken2023local,van2023integrating, kivits2022local}, our goal is to extend this approach to nonlinear interconnected systems.

Existing pHNN approaches often treat the interconnected system as a single unit, which limits their ability to take advantage of the compositional properties of port-Hamiltonian systems. Previous works on composite port-Hamiltonian systems assume direct access to state information or neglects the impact of noisy measurements. Additionally, these methods typically model subsystems in isolation before linking them together, failing to capture the dynamic interactions that occur when subsystems operate within a larger network \cite{ph_compositional}. Moreover, isolating a subsystem for experimentation is not always feasible. These limitations restrict the independent simulation and modification of subsystems, requiring re-identification of the entire system for any changes. 

\begin{figure}
    \centering
    \includegraphics[scale=0.6]{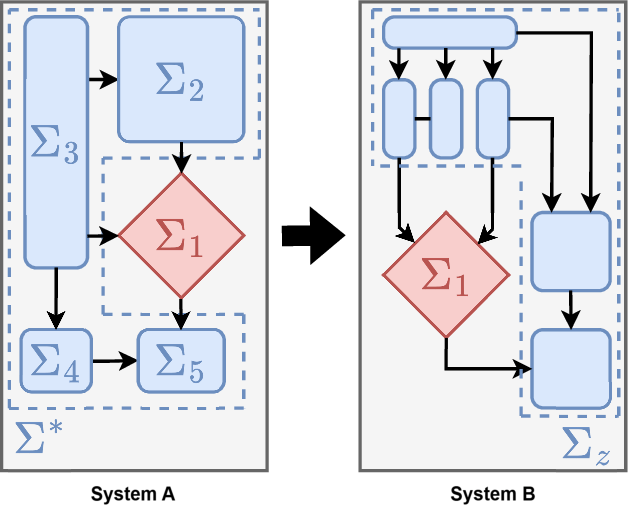}
    \caption{Motivation of the paper: learning the dynamics of a subsystem within one interconnected system such that it remains a valid model when the subsystem is part of another larger system. }
    \label{fig:motivation}
\end{figure}

To overcome these limitations, we introduce a new method for developing separable pHNN models that can handle measurement noise. This method allows for the independent modeling of subsystems while they operate within a larger system, thereby improving the compositionality and modularity of complex, nonlinear systems (see Fig.\ref{fig:motivation}).

The main contributions of this work are:
\begin{itemize}
\item \textit{A method to identify separable port-Hamiltonian models directly based on measured input-output data while the subsystem operates as part of a larger system.} Our method reconstructs the subsystem dynamics using only input-output data, avoiding the need for direct state measurements of the subsystems. 
\item \textit{Subsystem identification via noisy input-output data.} We handle measurement noise efficiently, using an OE noise setting, considering only input-output data.
\item \textit{Independent simulation and transferability of modeled subsystem dynamics.} Each subsystem can be simulated independently, allowing for model reuse and adaptation across different interconnected system configurations and physical domains.  
\end{itemize}

In this paper, in Section \ref{sec:Port-Hamiltonian representations}, we first introduce port-Hamiltonian representations and the composite port-Hamiltonian framework suitable for interconnected systems. Next, we present the proposed identification approach for pHNN networks in Section \ref{sec:Identification approach} and the corresponding modeling process. 
Finally, we demonstrate the effectiveness of the approach through simulation studies in Section \ref{sec:Simulation study}.

\section{Port-Hamiltonian representations}
\label{sec:Port-Hamiltonian representations}

\subsection{Input-State-Output equations}
\label{subsec:Input-State-Output equations}
As demonstrated in previous studies \cite{ph_theory, ph_book2}, port-Hamiltonian theory offers a physical modeling framework that is applicable to a wide range of systems. This framework not only describes these systems in interpretable terms, but also ensures properties such as passivity. The general input-state-output port-Hamiltonian equations are represented as 
\begin{subequations}\label{eq:ss-portHNN} 
\begin{align} 
    \dot{{x}}(t) &= [{J(x(t))-R(x(t))}] \nabla H(x(t)) + G(x(t))u(t), \label{subeq:ss-port_f}\\
    y(t) &= G^\intercal(x(t))\nabla H(x(t)),\label{subeq:ss-port_h}
\end{align}
\end{subequations} 
where $u(t) \in \mathbb{R}^{ n_\mathrm{u} }$, $x(t) \in \mathbb{R}^{n_\mathrm{x}}$ and $y(t)  \in \mathbb{R}^{n_\mathrm{y}}$ are the input, state and output vectors, respectively; with $n_\mathrm{y}=n_\mathrm{u}$ in terms \eqref{subeq:ss-port_h}. $H(x) \in \mathbb{R}$ is a scalar function that represents total energy of the system. $J(x) \in \mathbb{R}^{n_\mathrm{x} \times n_\mathrm{x}}$ is the skew-symmetric structure matrix, i.e., $ \forall x \in \mathbb{R}^{n_\mathrm{x}}: J(x)= -J^{\top}(x)$], $R(x) \in R^{n_{x} \times n_\mathrm{x}}$ is the symmetric positive semi-definite dissipation matrix, i.e,  $  \forall x \in \mathbb{R}^{n_\mathrm{x}}: R(x) = R^{\top}(x), x^\top R(x)x\geqslant 0$, and $G(x) \in R^{n_\mathrm{x} \times n_\mathrm{u}}$ is the external input matrix.

Port-Hamiltonian systems are inherently passive.  Passivity is a property that guarantees the system does not generate energy, which is essential for stability. Additionally, for cyclo-passivity, it is necessary that the Hamiltonian $H(x) $ is bounded from below. The differential dissipation inequality, which must be satisfied for passivity, is given by:
\begin{equation}\label{eq:passivity} 
    \frac{dH}{dt} = \frac{\partial H}{\partial {x}}^T J(x) \frac{\partial H}{\partial {x}} - \frac{\partial H}{\partial {x}}^T R(x) \frac{\partial H}{\partial {x}} + \frac{\partial H}{\partial {x}}^T G(x) u \leq y^T u.
\end{equation}
The inclusion of the output equation  \eqref{subeq:ss-port_h} is essential for maintaining passivity. Therefore, in the next section, when we introduce composite-port Hamiltonian systems, we will make sure to include the output equation.

\subsection{Composite port-Hamiltonian representation} \label{subsec: Composite port-Hamiltonian representation}
Port-Hamiltonian theory emphasizes port-based modeling, allowing systems to be intuitively connected to form larger interconnected systems. This approach facilitates the representation of these interconnected systems as networks of subsystems. A key property of port-Hamiltonian systems is their compositionality: when two port-Hamiltonian systems are interconnected through a power-preserving interconnection, they form a new composite port-Hamiltonian system \cite{schaft2021port}.

In the existing literature such as \cite{ph_compositional}, composite port-Hamiltonian representations generally do not consider the output equation, relying instead on direct state measurements of each subsystem to characterize the interconnected system behavior.  By including the output equation, our approach enables us to reconstruct the behavior of each subsystem solely on input-output measurement signals. This reduces the need for direct access to the internal states of each subsystem for later development of an identification algorithm, making the framework more practical for applications where state measurements are challenging or infeasible.

The objective of this section is to construct a representation of the interconnected port-Hamiltonian system by composing the terms introduced in \eqref{eq:ss-portHNN}. Let, $x_i(t) \in \mathbb{R}^{n_{x_i}}$, $u_i(t) \in \mathbb{R}^{n_{u_i}}$, and $y_i(t) \in \mathbb{R}^{n_{u_i}}$ represent the state, input, and output of the subsystem $i$, respectively. We define the state, input, and output of the interconnected system of $n$ subsystems as $x_c(t)=\mathrm{vec}(\{x_i(t)\}_{i=1}^{n}) \in \mathbb{R}^{n_{x_c}}$, $u_c(t)=\mathrm{vec}(\{u_i(t)\}_{i=1}^n) \in \mathbb{R}^{n_{u_c}}$with $y_c(t)$ similarly defined, where $n_{x_c}=n_{x_1}+...+n_{x_n}$, and $n_{u_c}=n_{u_1}+...+n_{u_n}$.

The dynamics of such an interconnected system are defined as \cite{ph_compositional}
\begin{subequations}\label{eq:c-portHNN}
\begin{align} \label{subeq:x_dot_c}
    \dot{x}_{c}(t) = & \left[ J_{c}(x_{c}(t)) - R_{c}(x_{c}(t)) + C \right]\nabla H_c(x_c(t)) \\
    & + G_{c}(x_{c}(t)) u_{c}(t), \nonumber\\
    \label{subeq:y_c}
    y_{c}(t) = & G_{c}^{\intercal}(x_{c}(t)) \nabla H_c(x_c(t));
\end{align}
\end{subequations}
here $J_c(x(t))$, $R_c(x(t))$, and $G_c(x(t))$ only contain block diagonal elements, i.e. $J_c(x_c(t)) = \mathrm{diag}(\{J_i(x_i(t))\}_{i=1}^{n})$ with a similar structure for $R_{c}(x(t))$, and $G_{c}(x(t))$. Within this structure, all of the interconnection information between subsystems is contained in $C$, which has zero entries on the (block-)diagonal and is skew-symmetric, preserving power in the interconnections. For now, the connection structure, $C$, is assumed to be known and independent of the state. Finally, note that the Hamiltonian for interconnected systems is simply the summation of the Hamiltonians of each subsystem, i.e. $H_{c}(x_{c}(t)) = \sum_{i=1}^{n} H_{i}(x_{i}(t))$.

In Equation \eqref{subeq:x_dot_c}, $J_c(x_c(t))$, $R_c (x_c(t))$ and $G_c (x_c(t))$ are all block diagonal matrices. This implies that the dynamics of each subsystem are determined by their own states and inputs, without direct dependence on the states of other subsystems. This means that the terms describing internal energy exchanges or dissipation within each subsystem are localized, allowing for the formulation of independent subsystem dynamics. The coupling between different subsystems is then entirely captured by the interconnection matrix $C$, which models how energy or power flows between subsystems.

\section{Identification approach} \label{sec:Identification approach}
\subsection{Considered data-generating system} 
Consider a composite port-Hamiltonian system that is described by \eqref{eq:c-portHNN} which can be written in terms of continuous time state-space equations as
\begin{subequations}\label{eq:system_class} 
    \begin{align} \label{subeq:state_time derivative system}
        \dot{x}_c (t) &= f({x_c}(t),{u_c}(t)) \\ \nonumber 
        &=[{J_c(x_c(t))-R_c(x_c(t))}+C] \nabla H_c(x_c(t)) + G_c(x_c(t)) u_{c}(t),
         \label{subeq:output function} \\
        y_{c0} (t) &= h({x_c}(t))\\ \nonumber 
        &= G_c^\intercal(x_c(t)) \nabla H_c(x_c(t)) , \\
        y_c(k) &= y_{c_0} (kT_s) + \varepsilon(k); \label{subeq:sampled measurements}
    \end{align}
\end{subequations}
where $T_s$ is the sampling time; here the outputs, $y_c(k)$, are assumed to contain measurement noise,  $\varepsilon(k)$, a zero-mean noise with finite variance. The collected input-output measurements are defined as
 \begin{equation}\nonumber 
 {\mathcal{D}_N=\{(y_c(k),u_c(k)\}_{k=0}^{N-1}}.
 \end{equation}
Here, a zero-order hold assumption is made for the input, implying that $u_c(t) = u_c(kT_{s}), \; \forall t \in [kTs, (k+1)Ts)$ with $k\in\mathbb{Z}$.

\begin{figure*}[h]
    \centering
    \includegraphics[scale=0.30]{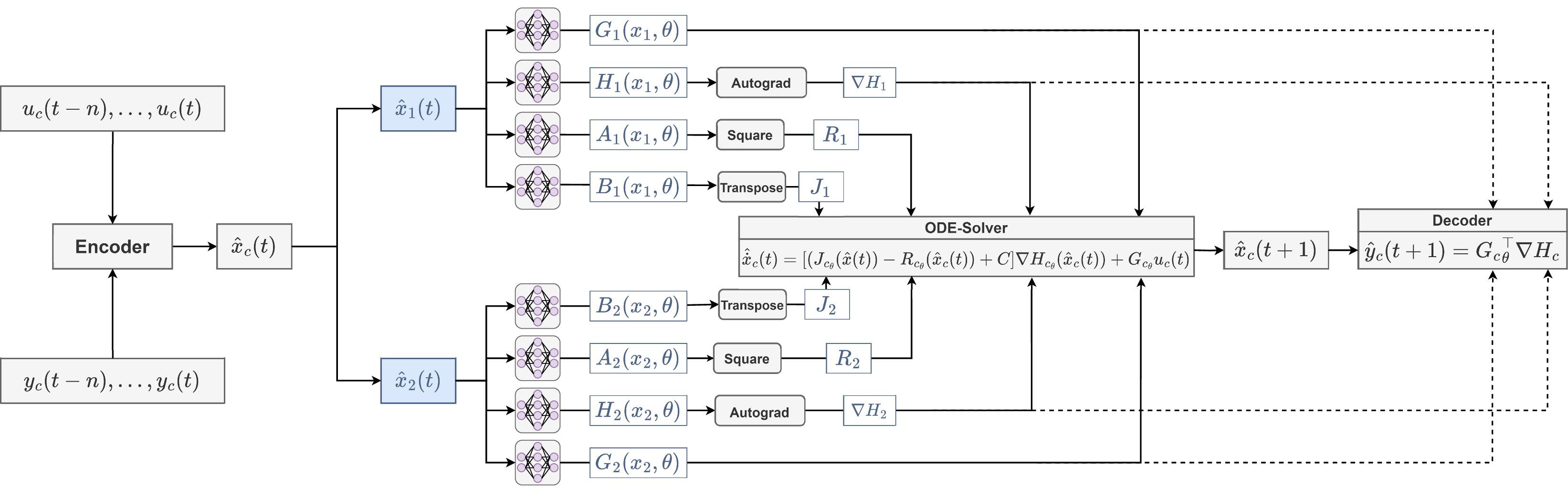}
    \caption{Schematic overview of the model structure for a composite system consisting of two subsystems. Note that the ODE-Solver step can be repeated for multiple timesteps to simulate further into the future. For the experiments in this paper, a neural network is used for the encoder, but different options are possible.}
    \label{fig:model-structure}
\end{figure*}

\subsection{Model structure} \label{subsec:model structure}
To model the dynamics of the considered systems \eqref{eq:system_class}, an \emph{output-error} (OE) model structure is selected. Recall Equation \eqref{eq:ss-portHNN} which can comprehensively represent the dynamics of interconnected systems, the parameterized model of the system can be defined as follows
\begin{subequations}
\label{eq:port_H_parametrized} 
    \begin{align}
    \label{subeq:f_theta}
    \dot{\hat{x}}_{c}(t) &= f_{c_{\theta}}(\hat{{x}}_{c}(t), u(t))   \\  \nonumber 
    &= [(J_{c_{\theta}}(\hat{x}(t)) - R_{c_{\theta}}(\hat{x}_c(t))+C]\nabla H_{c_{\theta}} (\hat{x}_c(t))\\  \nonumber 
    & \quad +  G_{c_\theta}(\hat{{x}}_c(t)) u_{c}(t), \\
    \hat{{y}}_c(t) &= h_{c_{\theta}}(\hat{{x}}_c(t)) = G_{c_\theta}^\intercal(\hat{{x}}_{c}(t)) \nabla H_{c_{\theta}} (\hat{x}_{c}(t)), \label{subeq:h_theta}\\ 
    \hat{y}_{c_{k}} &= \hat{y}_{c}(kT_{s}),
    \end{align}
\end{subequations}
where $\theta \in \Theta \subseteq \mathbb{R}^{n_\theta}$ is the vector of model parameters associated with $J_c, R_c, H_c$, and $G_c$ each formulated as structured neural networks. As noted in \eqref{eq:c-portHNN}, the matrix functions $J_{c_{\theta}}$, $R_{c_{\theta}}$, and $G_{c_{\theta}}$ are constructed from the individual subsystem matrices $J_{i_{\theta}}$, $R_{i_{\theta}}$, and $G_{i_{\theta}}$. The diagonal structure of these matrices allows each subsystem to be modeled independently, enabling us to treat the subsystem elements separately when forming the overall composed model. Also, since the total Hamiltonian $H_{c_{\theta}}$  is the sum of the Hamiltonians of each subsystem, $H_{i_{\theta}}$, we can model the Hamiltonian of each subsystem independently. 

We define \( R_{i_{\theta}} \) as the product of a matrix function \( A_{i_{\theta}} \) and its transpose, given by \( R_{i_{\theta}} = A_{i_{\theta}} A_{i_{\theta}}^\intercal \). This formulation ensures that \(R_{i_{\theta}} \) is a symmetric positive semi-definite matrix. Similarly, we formulate \( J_{i_{\theta}} \) as the difference between a matrix function \( B_{i_{\theta}} \) and its transpose, expressed as \( J_{i_{\theta}} = B_{i_{\theta}} - B_{i_{\theta}}^\intercal \). This formulation guarantees that \( J_{i_{\theta}} \) is skew-symmetric.  To impose a lower bound on $H_{i_{\theta}}$, and ensure cyclo-passivity, we use the exponential linear unit (ELU) activation function in the last layer of the neural network representing $H_{i_{\theta}}$, with a user-defined constant as the lower bound. All parameterized functions are implemented using multi-layer feedforward neural networks (MLPs).

\textbf{Remark 1.} The proposed model structure allows for flexibility in handling known and unknown elements of the interconnected system. Specifically, some subsystems of the system can be treated as known dynamics, while the remaining parts can be modeled as unknown. This enables us to selectively parameterize only the unknown aspects, simplifying the overall model and focusing computational resources on the areas where detailed identification is needed.

\textbf{Remark 2.} A sensible identification strategy focuses on modeling the target subsystems individually while grouping other subsystems into a single composite unit. For instance, if we are interested in modeling of subsystem $\Sigma_1$ (as shown in Fig.\ref{fig:motivation}), we provide detailed modeling for $\Sigma_1$ and treat the remaining components as one unified subsystem ($\Sigma^*$). This allows us to isolate and accurately represent $\Sigma_1$ while approximating the collective behavior of other subsystems, avoiding the need to model each subsystem individually. This approach preserves necessary interactions within the interconnected system while simplifying the overall structure.

\subsection{Identification of Composite pHNNs} \label{subsec: Identification of Composite pHNNs}
To identify the parameterized functions in the OE model introduced in Section \ref{subsec:model structure}, the standard approach is to minimize the simulation loss function, defined as:
\begin{equation}
    V_{\mathcal{D}_N}(\theta) = \frac{1}{N} \sum_{k=1}^{N} \left\| \hat{y}_{c_k} - y_{c_k} \right\|_2^2, \label{eq:loss_classic_sim} 
\end{equation}
where $\hat{y}_{c_{k}}$ and $y_{c_{k}}$ are the simulated and measured outputs, respectively. This minimization is subject to the system dynamics defined by the parameterized port-Hamiltonian model in \eqref{eq:port_H_parametrized}. 

In this paper, we adopt the SUBNET approach \cite{MS_subnet}, which breaks the data into multiple shorter subsections each of length $T$ (the truncation length), rather than using the entire dataset at once. This strategy makes the optimization process more manageable and robust. For each subsection, an initial state is estimated using an encoder function:
\begin{equation}
\hat{x}_{t|t} = \psi_\eta (y_{t-n}^{t-1}, {u}_{t-n}^{t-1}),
\end{equation}
where $\psi_{\eta}$ is an encoder parameterized by $\eta$, with $n$ representing the length of the input-output data used for state estimation.  Here, ${u}_{t-n}^{t-1} = \begin{bmatrix} u_{t-n}^\intercal  \cdots  u_{t-1}^\intercal \end{bmatrix} ^\intercal$; with $y_{t-n}^{t-1}$ defined in the same way. 

The loss function introduced in \eqref{eq:loss_classic_sim}  is then calculated across these subsections as: \vspace{1mm}
\begin{subequations}\label{eq:SUBNET_pHNN}
\begin{align} 
    V^{\text{sub}}_{\mathcal{D}_N}(\theta,\eta) &= \frac{1}{\tau} \sum_{t=n+1}^{N-T+1}\sum_{k=0}^{T-1} \left\|y_{t+k}- \hat{y}_{t+k|t}\right\|_2^2, \label{subeq:loss_theta}\\  
    \intertext{\vspace{-4mm} subject to:} 
    \hat{x}_{t|t} &= \psi_\eta (y_{t-n}^{t-1}, u_{t-n}^{t-1}), \label{subeq:x_theta}\\
    \hat{{x}}_{c_{t+k+1|t}} &= \text{ODE-solver} [f_{c_{\theta}}(\hat{{x}}_{c_{t+k|t}}, {u}_{t+k})], \label{subeq:x_k+1_theta}\\
    \hat{y}_{c_{t+k|t}} &= h_{c_{\theta}}(\hat{{x}}_{c_{t+k|t}})    \label{subeq:y_theta}
\end{align}
\end{subequations}
where $\tau=(N-T+1)T$, and functions $f_{c_{\theta}}$, and $h_{c_{\theta}}$ are as defined in equations \eqref{subeq:f_theta}, and \eqref{subeq:h_theta}, respectively. We employed the fourth-order Runge-Kutta method (RK4) \cite{RK4} as our ODE-solver, using a time-step of $T_{s}$. The $(|)$ represents the current time index given the initial state of the subsection.

Overall, the model architecture consists of three main components:
\begin{enumerate}
    \item Encoder ($\psi_\eta$): Estimates the initial state for each subsection based on past input-output data.
    \item State Evolution ($f_{c_{\theta}}$): Predicts future states using an ODE solver.
    \item Decoder ($h_{c_{\theta}}$):  Computes the output of the system from the predicted states.
\end{enumerate}

By repeating these steps over each timestep within the subsections, the model simulates the behavior of the system across the entire dataset. The details of this process are outlined in Fig. \ref{fig:model-structure}. This approach optimizes the parameters of the encoder, state evolution, and decoder while integrating the port-Hamiltonian framework.

\section{Simulation study} \label{sec:Simulation study}
To provide insight into the performance of the composite port-Hamiltonian approach, this section studies two numerical examples and their validation.

\subsection{Data-generating system}
\begin{figure}[h]
   \centering
   \includegraphics[scale=0.3]{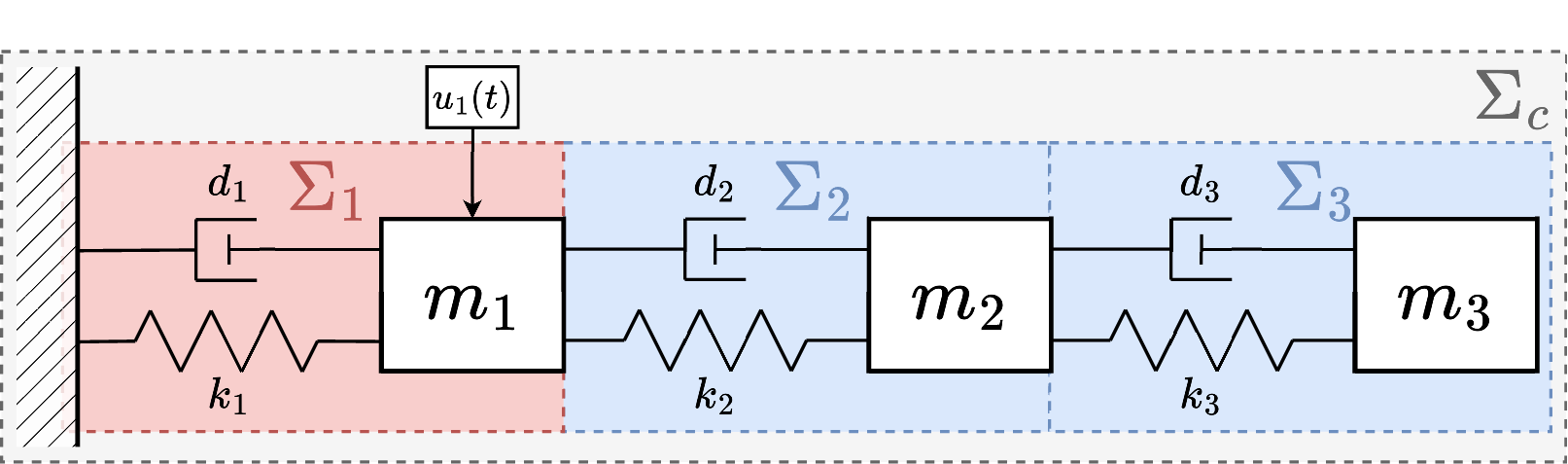}
   \caption{Schematic view of the three coupled mass-spring-dampers. For this system, the input is given as a force $u_{1}(t)$, the states are the displacements, ${q}(t)$, and momenta, ${p}(t)$, of the masses, while the outputs are the velocities of the masses $\dot{{q}}(t)$. Note that this can be viewed as a single composite system, or as three interacting subsystems.}
   \label{fig:3-MSD}
\end{figure}

Consider a system of three coupled mass-spring-dampers (MSDs) with an external force applied to the first mass as illustrated in Fig. \ref{fig:3-MSD}. The system is governed by nonlinear cubic damping dynamics, described by the equation \vspace{1mm}
\begin{equation} \label{eq:3MSD}
    M \ddot{q}(t) + D \dot{q}^3(t) + Kq(t) = u(t),
\end{equation}
where \( M \), \( D \), and \( K \) are the mass, damping, and spring matrices respectively, while $q(t) \in \mathbb{R}^{3}$ represents the mass displacements. Parameters for all three MSDs are identical and set as \( k_{i} = 1 \), \( m_{i} = 2 \), and \( d_{i} = 0.5 \).

A multisine input, composed of 40 frequencies, is applied as an external force to the first mass to excite that subsystem, defined as \vspace{1mm} 
\begin{equation}
    u_1(t) = \sum_{i=1}^{40} \sin(2 \pi i f_0 t + \phi_i),
\end{equation}
with \( f_0 = 0.1 \) and random phases \( \phi_i \) uniformly sampled from \([0, 2\pi)\). The system is simulated for 250 seconds, with the velocity of each mass sampled at 10 Hz, resulting in a dataset of 2500 samples. Uniform white noise is added to the output signals, producing a \emph{signal-to-noise ratio} (SNR) of approximately 20 dB. We measure the output of all subsystems and excite one subsystem. In a LTI setting, it is shown that this leads to an identifiable network \cite{SI_lizan_thesis, SI_lizan_CH10}. The identifiability of nonlinear PH networks remains the topic of future work. The resulting dataset consisting of input-output data is used for identification, with the internal states associated with the system remaining hidden.

\subsection{Data-set and training} \label{subsec:sim_training}
\begin{figure}
    \centering
    \includegraphics[width=\linewidth]{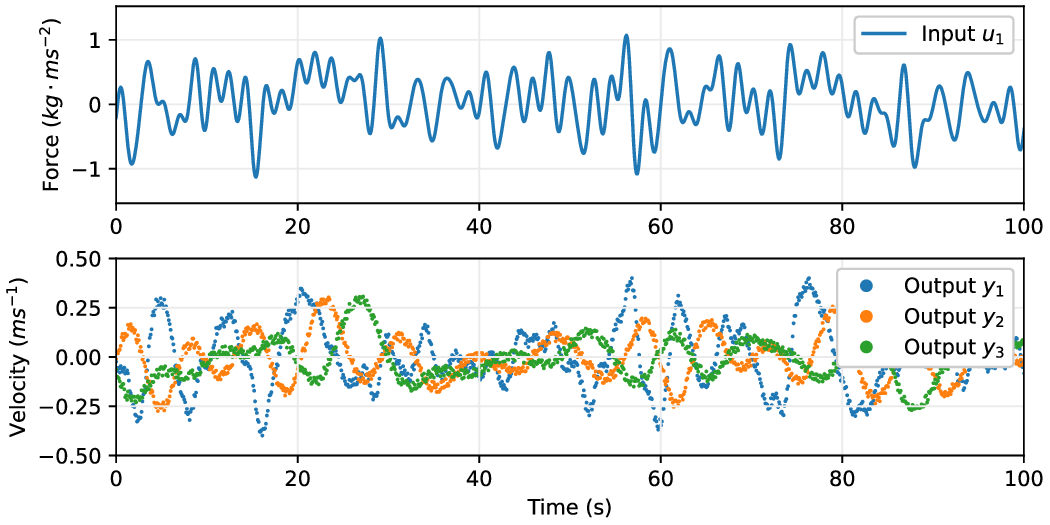}
    \caption{Example of the three MSD system behavior for the first 100 seconds. The first subplot shows the external force that is taken as an input, $u_1$. The second subplot shows the corresponding output measurements of the velocities of the three masses, $y$.}
    \label{fig:dataset-example}
\end{figure}

We generated eight sets of input-output data. Out of these, five sets were used for training the model, two sets for validation, and one set for testing. Fig. \ref{fig:dataset-example} illustrates the input and output of the system for the first 100 seconds.

The training process involved applying the algorithm to the training datasets to optimize the neural network parameters by minimizing the loss function \eqref{eq:loss_classic_sim}. The encoder used in this process was a residual neural network with two hidden layers, each containing 64 nodes. The matrices $J_{i_\theta}$, $R_{i_\theta}$, and $H_{i_\theta}$ were implemented using MLPs with a single hidden layer of 16 nodes. All networks used the hyperbolic tangent as their activation function. Since $G_{c\theta}$ is a constant matrix as it is assumed to be state-independent, its matrix elements are optimized directly instead. To improve convergence, this matrix is initialized as identity.

All neural networks and parameters were trained simultaneously using the ADAM optimizer \cite{AP_adam}, with a learning rate of $10^{-3}$, over a fixed duration of 1000 epochs.  Although early stopping was incorporated into the process, it was not required, as the training loss plateaued without signs of overfitting before reaching the final epoch. This approach ensured that the neural network was well-trained and optimized for accurate performance.

\subsection{Results} 
\begin{figure}
    \centering
    \includegraphics[width=\linewidth]{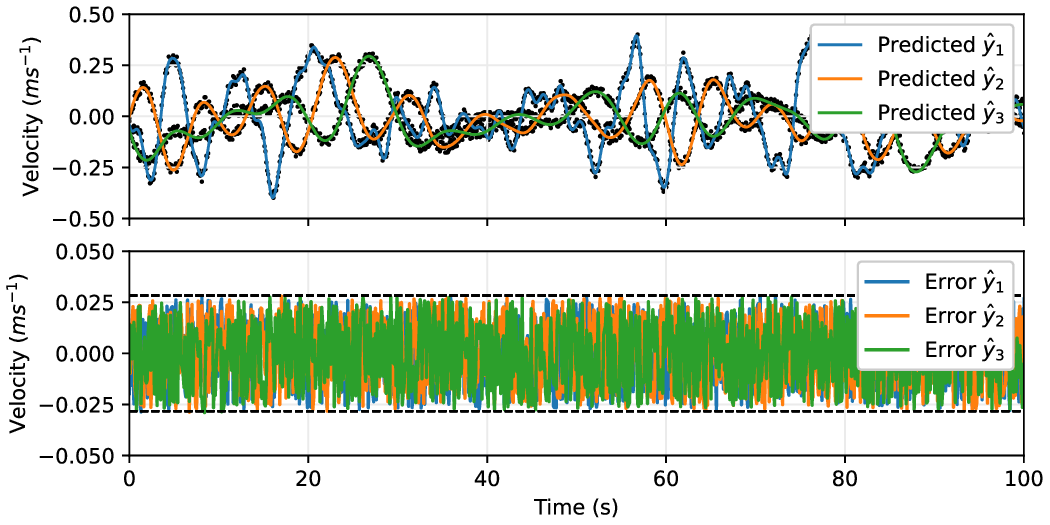}
    \caption{Simulation of the identified model for the first 100 seconds. In the first subplot, the dots represent the sampled true output values, while the solid lines indicate the simulation response of the model. The second subplot shows the error between the simulated and measured outputs. The amplitude of the noise is indicated with the dashed lines.}
    \label{fig:training-qualitative}
\end{figure}

Fig. \ref{fig:training-qualitative} presents the qualitative results of the trained model on the test dataset. In this figure, the simulated outputs of the model are compared with the measured output data from the data-generating system. At first glance, it is evident that the model performs well, as the simulated outputs closely follow the trend of the measured outputs. Looking at the error plot supports this conclusion, as the difference between the measurements and the simulated responses of the model is approximately at the noise floor level. 

\subsection{Multi-physics system} \label{subsec:multi-physics}
\begin{figure}[h]
   \centering
   \includegraphics[scale=0.35]{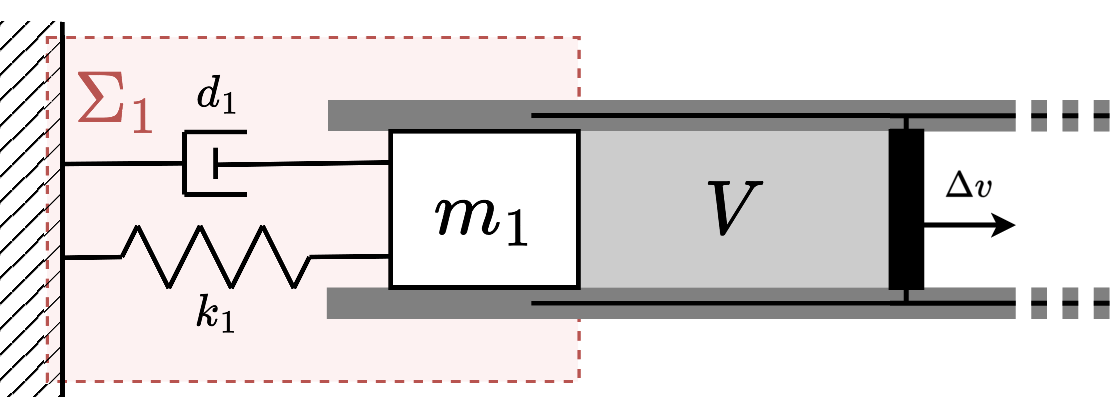}
   \caption{Schematic overview of the multi-physics system. The MSD is placed in a frictionless cylinder that is kept at a constant temperature, with an ideal gas reservoir on one end. The states associated with the PH model of the system are the displacement and momentum of the mass, $q$ and $p$, as well as the volume of the reservoir, $V$. The outputs are the velocity of the mass, $\dot{q}$, and the reservoir pressure, $P_{v}$.}
   \vspace{-0.5cm}
   \label{fig:multi-physics}
\end{figure}

To validate that the learned subsystem dynamics are indeed transferable, another experiment is needed. In this experiment, the trained subsystem networks, $J_{1\theta}(x_{1})$, $R_{1\theta}(x_{1})$, $G_{1\theta}$ and $H_{1\theta}(x_{1})$, are extracted and connected to the known dynamics of another system. This results in the following construction,
\begin{equation}
    \begin{aligned}
    \begin{bmatrix}
        \dot{x}_{1} \\
        \dot{x}_{z}
    \end{bmatrix}
    = &
    \begin{bmatrix}
        J_{1\theta}(x_{1})-R_{1\theta}(x_{1}) & C_{1z} \\
        C_{z1} & J_{z}(x_{z})-R_{z}(x_{z})
    \end{bmatrix}
    \begin{bmatrix}
        \nabla H_{1\theta}(x_{1}) \\
        \nabla H_{z}(x_{z})
    \end{bmatrix} \\
    & +
    \begin{bmatrix}
        G_{1\theta} & 0 \\
        0 & G_{z}
    \end{bmatrix}
    \begin{bmatrix}
        u_{1} \\
        u_{z} 
    \end{bmatrix},
    \end{aligned}
\end{equation}
where $J_{z}$, $R_{z}$, $G_{z}$ and $H_{z}$ represent the known dynamics of the new system. We chose to connect the first MSD with an ideal gas reservoir representing the new system in order to simultaneously validate the cross-domain viability of the proposed approach. The setup is shown in Figure \ref{fig:multi-physics}, in which the input is a slider at the end of the reservoir, which can adjust the volume and, consequently, induce changes in the pressure of the gas. It can be represented in pH form as
\begin{equation}
    \begin{bmatrix}
        \dot{q} \\
        \dot{p} \\
        \dot{V} \\
    \end{bmatrix}
    =
    \begin{bmatrix}
        0 & 1 & 0 \\
        -1 & -d_1 \dot{q}^{2} & -A \\
        0 & -A & 0 \\
    \end{bmatrix}
    \begin{bmatrix}
        k_1q \\
        \frac{p}{m_1} \\
        \frac{\gamma}{V}
    \end{bmatrix}
    +
    \mathbf{I}
    \begin{bmatrix}
        0 \\
        0 \\
        \Delta V \\
    \end{bmatrix},
\end{equation}
where $V$ is the volume of the gas reservoir and $\Delta V$  is the change in volume caused by the slider. The cross-section of the interior of the cylinder and the dimensionless gas constant are taken as constants and set as $A=5$, $\gamma=1$, respectively.

\begin{figure*}[h]
   \centering
   \includegraphics[width=\linewidth]{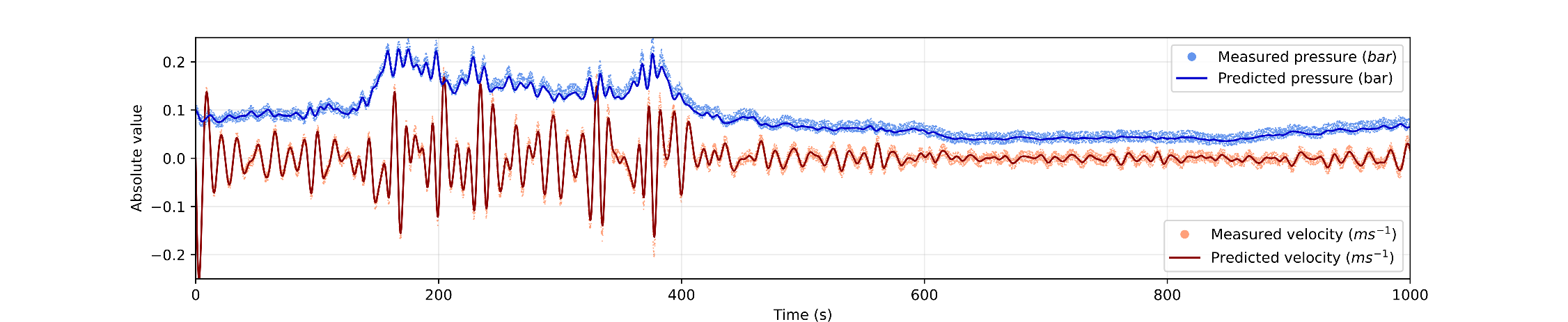}
   \caption{Simulation results of multi-physics system. The blue and red dots represent the measured values for the pressure of the reservoir and the velocity of the mass respectively. The solid lines indicate the corresponding simulated values over the course of the simulation.}
   \label{fig:exp-3}
\end{figure*}

Figure \ref{fig:exp-3} shows the modeled output compared to the noisy system outputs. It can be seen that the model follows the system outputs closely, indicating that the learned dynamics have been transferred successfully. 

\section{Conclusions}
This paper presents a novel approach for identifying and modeling individual subsystems within interconnected port-Hamiltonian systems using input-output measurements. The proposed method employs the port-Hamiltonian Neural Network (pHNN) to model each subsystem independently, enabling their separate simulation. Significantly, this method transforms noisy input-output data into interpretable port-Hamiltonian models without requiring knowledge of internal states. Its effectiveness is demonstrated through numerical studies, where the learned subsystem dynamics from one interconnected system were successfully transferred to another multi-physics model. This highlights the potential of using pHNNs in various engineering applications where physical interpretability and subsystem-level modeling are essential.




\bibliographystyle{ieeetr}
\bibliography{reference}

\end{document}